\documentclass[mancolor,aps,preprint]{revtex4}
\usepackage{color,graphicx}
\begin{document}

\begin{center}

{\large\bf Hall-effect evolution across a heavy-fermion quantum critical
point}\\[0.5cm]

S.~Paschen$^{\ast}$, T.~L{\"u}hmann$^{\ast}$, S.~Wirth$^{\ast}$,
P.~Gegenwart$^{\ast}$, O.~Trovarelli$^{\ast}$, C.~Geibel$^{\ast}$,
F.~Steglich$^{\ast}$, P.~Coleman$^{\dag}$ \& Q.~Si$^{\ddag}$\\

{\em $^{\ast}$Max Planck Institute for Chemical Physics of Solids,
N{\"o}thnitzer Str.~40, D - 01187~Dresden, Germany}\\

{\em $^{\dag}$Center for Materials Theory, Department of Physics and
Astronomy, Rutgers University, Piscataway, New Jersey 08855, USA}\\

{\em $^{\ddag}$Department of Physics and Astronomy, Rice University, Houston,
Texas 77005-1892, USA}

\end{center}

\vspace{0.5cm}
{\bf A quantum critical point (QCP) develops in a material at absolute zero
when a new form of order smoothly emerges in its ground state. QCPs are of
great current interest because of their singular ability to influence the
finite temperature properties of materials. Recently, heavy-fermion metals have
played a key role in the study of antiferromagnetic QCPs. To accommodate the
heavy electrons, the Fermi surface of the heavy-fermion paramagnet is larger
than that of an antiferromagnet\cite{Mar82.1,Ful88.1,Osh00.1}. An important
unsolved question concerns whether the Fermi surface transformation at the QCP
develops gradually, as expected if the magnetism is of spin density wave (SDW)
type\cite{Her76.1,Mil93.1}, or suddenly as expected if the heavy electrons are
abruptly localized by magnetism\cite{Sch00.1,Col01.1,Si01.1}. Here we report
measurements of the low-temperature Hall coefficient ($R_H$) - a measure of the
Fermi surface volume - in the heavy-fermion metal YbRh$_2$Si$_2$ upon
field-tuning it from an antiferromagnetic to a paramagnetic state. $R_H$ 
undergoes an increasingly rapid change near the QCP as the temperature is
lowered, extrapolating to a sudden jump in the zero temperature limit. We
interpret these results in terms of a collapse of the large Fermi surface and
of the heavy-fermion state itself precisely at the QCP.}

\newpage

The compound YbRh$_2$Si$_2$ investigated here appears to be one of the best
suited heavy-fermion metals known to date to study the evolution of the Hall
effect across a QCP. Magnetic susceptibility and specific heat indicate that it
orders antiferromagnetically via a second-order phase transition at very low
temperatures ($T_N =70$~mK)\cite{Tro00.2}. The antiferromagnetic nature of the
transition is supported by NMR data\cite{Ish02.1}. Neutron scattering
experiments to directly detect the magnetic order are, presumably due to the
smallness of the ordered moment\cite{Ish03.2}, not available to date. The
N{\'e}el temperature $T_{N}$ is continuously suppressed down to the lowest
experimentally accessed temperatures by application of a small magnetic field
($B_{1c} \approx 0.7$~T for a field along the magnetically hard $c$-axis,
$B_{2c} \approx 60$~mT for a field within the easy tetragonal
plane)\cite{Geg02.1}. In addition, isothermal magnetostriction measurements
indicate that the transition remains of second order down to at least 15~mK
(ref.\ 13).
Although a change from second to first order at even lower temperatures can, of
course, not be strictly ruled out the non-Fermi liquid behaviour observed for
three decades of temperature (10~mK$ < T < 10$~K)\cite{Geg02.1} is best
described within a quantum critical picture. The use of tiny fields permits
one to reversibly access the QCP without the introduction of additional
disorder and without altering the character of the underlying zero-field
transition\cite{Cus03.1}. Moreover, unlike for several other heavy-fermion
compounds\cite{Ste01.1} (and the high-$T_c$ superconductors), the QCP is not
hidden by superconductivity. This is in spite of the high quality of the
YbRh$_2$Si$_2$ single crystals investigated here (residual resistivities of
$\approx 1$~$\mu\Omega$cm,
ref.\ 12).
The scaling analysis of the {\em thermodynamic} and {\em dynamical} properties
(specific heat, magnetic susceptibility, electrical resistivity)
suggests\cite{Geg02.1} that the field-induced QCP in YbRh$_2$Si$_2$ is of
local\cite{Sch00.1,Col01.1,Si01.1} rather than of
itinerant\cite{Her76.1,Mil93.1} type, similar to the doping-induced QCP in
CeCu$_{6-x}$Au$_x$
(ref.\ 6).
Hall-effect measurements may be used to access a {\em static electronic}
property, namely the Fermi surface volume, for which clear-cut theoretical
predictions exist for different types of QCP\cite{Col01.1,Si01.1,Sen04.1}. The
study presented here is the first systematic Hall-effect measurement at a
heavy-fermion QCP.

The Hall effect, usually a rather complex quantity, appears to be surprisingly
simple here, both in vanishingly small and in finite magnetic fields. Outside
the quantum critical region the Hall resistivity is linear in field resembling
the behaviour of simple metals.  Furthermore, our analysis of the
temperature-dependent Hall coefficient in terms of the anomalous Hall effect
(Fig.\,\ref{HallT}a, Methods, and 
refs.\ 17 and 18)
reveals that the low-temperature (below about 1~K) Hall coefficient is
dominated by its {\em normal} contribution. These features imply that the
low-temperature Hall coefficient can be used, to a good approximation, as a
measure of the Fermi surface volume. In the absence of photoemission and de
Haas-van Alphen studies (the latter presumably never being available because of
the very low critical magnetic field) as well as of electronic bandstructure
calculations this is, so far, the only information on the Fermi surface volume
of YbRh$_2$Si$_2$. At zero magnetic field, the data measured at the lowest
temperatures tend to saturate at the value of the normal Hall coefficient
extracted from the data between 7~K and room temperature (Fig.\,\ref{HallT}a).
This indicates that, at $B = 0$, the Fermi surface volume is the same at the
lowest temperatures as it is at high temperatures. Thus, even though there is
evidence for the onset of Kondo screening at approximately 20~K
(refs.\ 9 and 12)
and for surprisingly large effective quasiparticle masses in the
antiferromagnetically ordered state close to the QCP\cite{Geg02.1}, the local
moments do, at the lowest temperatures and at $B = 0$, not appear to be
incorporated into the Fermi surface. In the {\em static}
sense\cite{Mar82.1,Ful88.1,Osh00.1}, YbRh$_2$Si$_2$ may, in its unconventional
antiferromagnetic state at $B < B_c$, therefore not be classified as a
``heavy-fermion'' metal. 

In the intermediate temperature range, between approximately 70~mK and 7~K,
there is an additional contribution $\Delta R_H$ which is not due to the
anomalous Hall effect (Fig.\,\ref{HallT}a). In the main part of
Fig.\,\ref{HallT}b we show that, between 0.7 and 5~K, the cotangent of the Hall
angle is linear in $T^2$ (while $\Delta\rho \propto T$), indicating that the
longitudinal and transverse scattering rates are different\cite{And91.1}. This
type of behaviour is well known in high-$T_c$ cuprates where it has been taken
as evidence for spin-charge separation\cite{And91.1}. Note, however, that for
YbRh$_2$Si$_2$ the temperature range where this relation holds (inset of
Fig.\,\ref{HallT}b), is narrower than the one where the non-Fermi liquid (NFL)
behaviour $\Delta\rho \propto T$ is observed (from 100~mK to almost 10~K,
ref.\ 12),
even if YbRh$_2$Si$_2$ is field-tuned to quantum criticality (green squares in
inset). The same may hold true for CeCoIn$_5$
(ref.\ 20). 

In our field-dependent Hall-effect measurements on YbRh$_{2}$Si$_{2}$ the
magnetic field plays dual roles, as both a ``tuning'' and a ``probe'' field. On
the one hand, the coupling  between the field and the Yb$^{3+}$ moments tends
to align the latter: it is this Zeeman-like  coupling which tunes the ground
state of the material, ultimately suppressing the antiferromagnetism and
creating the QCP. On the other hand, the magnetic field also generates a weak
Lorentz force on the underlying electrons which produces the Hall response. The
weak orbital  coupling responsible for the Lorentz force does not appreciably
change the ground state so that, to a good approximation, we can discuss the
two couplings independently. The single crystals of YbRh$_{2}$Si$_{2}$ are thin
platelets oriented along the $ab$-plane and practical Hall-effect measurements
require a current and Hall voltage lying in this plane. This allows for two
distinct types of experiment, ``transverse tuning'' where the tuning field
$B_{1}$ is parallel to the $c$-axis, perpendicular to the current, and
``longitudinal tuning'' where the tuning field $B_{2}$ lies parallel to the
current in the basal plane (cf.\ schematics in Figs.\,\ref{HallB}a and b). The
longitudinal field $B_{2}$ produces essentially no Hall response (see
Supplementary Methods 1) and serves only to tune the state: a separate, crossed
probe field $\delta B_1$ along the $c$-axis is required to measure the Hall
response. In this longitudinal (crossed-field) experiment, the Hall resistivity
$\rho_H$ is a direct measure of the field-tuned (linear-response) Hall
coefficient $R_{H}(B_{2})$
\begin{equation}
R_{H}(B_{2}) \equiv \lim_{B_1\rightarrow 0}\rho_H(B_2,B_1)/B_1.
\label{linear-response}
\end{equation}
In the transverse (single-field) case, on the other hand, the magnetic field
simultaneously  tunes the state and probes the Hall response and the {\em
differential} Hall coefficient $\tilde{R}_{H}(B_1)$ is 
\begin{equation}
\tilde{R}_{H}(B_1) \equiv \frac{d\rho_H(B_1)}{dB_1} 
=\left [ \frac{\partial\rho_H(B_1)}{\partial B_1}
 \right ]_{\rm orb}
+ \left [ \frac{\partial\rho_H(B_1)}{\partial B_1} \right ]_{\rm zeeman}
=R_H(B_1) + 
\left [ \frac{\partial\rho_H(B_1)}{\partial B_1} \right ]_{\rm zeeman}.
\label{differential-response}
\end{equation}
The orbital (``probing'') contribution is, according to the Kubo formalism,
just the generalized definition of a Hall coefficient (see Supplementary
Methods 2). The Zeeman (``tuning'') term is not related to a readily measurable
linear-response quantity. 

We first discuss the results of the single-field experiment.
Figure\,\ref{HallB}a displays several representative isotherms of the Hall
resistivity $\rho_H$, corrected for its anomalous contribution $\rho_{H,a}(B)$
(see Methods), vs $B_1$.  $\rho_H - \rho_{H,a}$ shows a linear low-$B_1$
behaviour with larger and a linear high-$B_1$ behaviour with smaller slope. The
crossover between the two regimes broadens and shifts to higher $B_1$ with
increasing temperature. For a quantitative analysis of the data we choose
$\tilde{R}_{H}(B)= R^{\infty}_{H}- (R^{\infty}_{H} - R^{0}_{H})\gamma(B)$ as a
fitting function, where $R^{0}_{H}$ is the zero-field Hall coefficient and
$R^{\infty}_H$ is the asymptotic differential Hall coefficient at large fields.
$\gamma(B)$ is a crossover function that changes from unity at low fields to
zero at large fields, which we parameterize as $\gamma(B) = 1/[1+(B/B_0)^p]$.
Here, $B_{0}$ is the crossover field and $p$ determines the sharpness of the
transition, which has a width $\Gamma\sim B_{0}/p$ when $p$ is large. For
$p\rightarrow\infty$, $\int{\tilde{R}_{H}(B)dB}$ has a sharp kink at  $B =
B_0$, corresponding to a step in $\tilde{R}_{H}(B)$ itself. The fits to the
data are shown as solid lines in Fig.\,\ref{HallB}a. For one temperature the
derivative of the fit, corresponding to $\tilde{R}_{H}(B_1)$, is shown as well.
The crossover fields $B_0$ obtained from these fits are included as red dots in
the temperature-field ($T-B$) phase diagram of YbRh$_2$Si$_2$
(Fig.\,\ref{PD}a). A linear fit to these points (dashed red line in
Fig.\,\ref{PD}a denoted $T_{Hall}$) extrapolates at zero temperature to the
critical field $B_{1c} \approx 0.7$~T for the disappearance of
antiferromagnetic order and the QCP. Thus, the crossover is directly related to
the QCP. The sharpness of the crossover is best quantified by the full width at
half maximum (FWHM) of $d\tilde{R}_{H}/dB_1$, which represents the change of
slope of $\rho_H(B_1)$. The temperature dependence of the FWHM values is well
described by a {\em pure} power law, ${\mathrm{FWHM}}\propto T^a$,
$a=0.5\pm0.1$ (inset of Fig.\,\ref{PD}a), suggesting  that, at zero
temperature, the crossover is infinitely sharp (FWHM~=~0).

To understand the origin of this sharp feature, we now turn to the
crossed-field measurement of the linear-response Hall coefficient,
Eq.~(\ref{linear-response}).  The inset of Fig.\,\ref{HallB}b displays
$\rho_H(B_1)$ curves taken at 65~mK for different values of the longitudinal
tuning field $B_2$. With increasing $B_2$ the linear-response Hall coefficient
$R_H$ decreases. For a quantitative analysis we fit, as above,
$\int{\tilde{R}_H(B)dB}$ to the $\rho_H(B_1)$ data (solid lines in the inset of
Fig.\,\ref{HallB}b). As opposed to the single-field experiment, $R^{0}_{H} =
R_H$ is now the only parameter to consider. $R_H$, normalized to its value at
the crossover field $B_0$, is plotted in the main panel of Fig.\,\ref{HallB}b
as a function of the normalized tuning field $B_2/B_0$. Data obtained in the
same way at 45, 75, and 93~mK are included as well. $R_H$ decreases as a
function of $B_2$ by a factor of $\approx 1.5$. In a simple one band model this
corresponds to an increase in the charge carrier concentration from $\approx 2$
to $\approx 3$ holes per YbRh$_2$Si$_2$ formula unit. The crossover sharpens up
as the temperature is lowered. For a quantitative analysis we may now fit the
crossover form $R_H(B)= R^{\infty}_{H}-(R^{\infty}_{H} - R^{0}_{H})\gamma(B)$
to the $R_H(B_2)$ data (solid curves in main panel of Fig.\,\ref{HallB}b). The
$R^{\infty}_{H}$ values obtained for these four temperatures are included as
green triangles in the main part of Fig.\,\ref{HallT}a, showing that the Hall
coefficient in the field-induced Landau Fermi liquid (LFL) state (cf.\
Fig.\,\ref{PD}a) at very low temperatures is substantially smaller than in the
$B=0$ antiferromagnetically ordered state.  The $11 B_0$ and FWHM values
obtained from the above fits are included as green dots in Fig.\,\ref{PD}a and
its inset. The factor of 11 accounts for the fact that the tuning field $B_2$
is applied in the easy tetragonal plane of YbRh$_2$Si$_2$ where, due to the
magnetic anisotropy, the action of a magnetic field is known to be $\approx 11$
times as strong as along the magnetically hard $c$-axis\cite{Geg02.1}. For both
quantities the green and red data points agree within the error bars. Thus, the
linear Hall response $R_{H}(B_2)$ of the crossed-field measurement and the
differential Hall response $\tilde{R}_H(B_1)$ of the single-field measurement
can be described by the same functional form and the respective crossover
positions and crossover widths agree quantitatively. This experimental finding
suggests that the second term on the right hand side of
Eq.~(\ref{differential-response}) plays a minor role, at least in the
experimentally accessed part of the $T-B$ phase diagram. Therefore, here the
single-field experiment appears to probe essentially the same (linear-response)
Hall coefficient as the crossed-field experiment. However, there is a
quantitative difference in the jump heights of $\tilde{R}_H(B_1)$ and
$R_{H}(B_2)$ which probably reflects the anisotropies in the evolution of the
electronic bandstructure under transverse and longitudinal
field-tuning\cite{Cus04.1}, amplified by the likely presence of a multisheeted,
anisotropic Fermi surface.

The phase diagram in the magnetic field-temperature parameter space can now be
illustrated by a 3D representation of $d\gamma(B)/dB$ (Fig.\,\ref{PD}b).
$\gamma(B)$ is calculated at arbitrary temperatures from the linear $B_0$ vs
$T$ fit (dashed red line in Fig.\,\ref{PD}a) and a power law fit (not shown) to
the $p(T)$ data obtained from the fits to $\rho_H(B_1)$ (Fig.\,\ref{HallB}a)
and to $R_H(B_2)$ (main panel of Fig.\,\ref{HallB}b). With decreasing
temperature, the $d\gamma(B)/dB$ curves sharpen up and their crossover position
$B_0$, designated by drop lines, shifts to lower fields such that, at zero
temperature, a $\delta$-function (dashed line in $T=0$ plane in
Fig.\,\ref{PD}b) is situated at the QCP.

Thus, the extrapolation of our finite temperature data to {\em zero
temperature} indicates the presence of a finite discontinuity (``jump'') in the
Hall coefficient at the QCP, even though the change in the magnetic order
parameter is infinitesimal\cite{Ish03.2}. By contrast, in an itinerant SDW
scenario, the Fermi surface is expected\cite{Col01.1} to fold over at the QCP;
the Hall coefficient is then continuous across the QCP, evolving gradually with
the size of the antiferromagnetic order parameter, as is indeed observed
experimentally\cite{Lee04.1}. (For a more quantitative comparison, see caption
of Fig.\,\ref{HallB}b.) Our results hint at a sudden reconstruction of the
Fermi surface at the QCP, corresponding to the sudden loss of ``mobile'' $4f$
electrons\cite{Col01.1,Si01.1,Sen04.1}. Loosely speaking, the volume of the
Fermi surface has changed discontinuously. Here of course, the concept of Fermi
surface volume needs to be treated with some care, for antiferromagnetism
breaks translational symmetry. YbRh$_2$Si$_2$ may well be an easy-plane
incommensurate antiferromagnet, and for this class of antiferromagnet, to
linear order in the magnetic order parameter, the Fermi surface volume is
well-defined in the paramagnetic unit cell. From our data we infer that the
antiferromagnetic ground state has a ``small'' Fermi surface which is the same
as the one extracted from the high-temperature Hall effect data (main panel of
Fig.\,\ref{HallT}a) while the paramagnetic ground state has a ``large'' Fermi
surface which presumably counts the new heavy-fermion states injected by the
local moments.

The crossover line $T_{Hall}(B)$ (Fig.\,\ref{PD}) is then interpreted as the
{\em finite temperature} signature of the ``jump'' in the Fermi surface volume.
It delineates the position at which a new large Fermi surface emerges in the
incoherent electron fluid. It is interesting that this pre-cursor to heavy
quasiparticle formation takes place at temperatures well above the temperature
$T^{\ast}$ (dashed black curve in Fig.\,\ref{PD}a) below which the coherent LFL
develops. The existence of a large Fermi surface in the absence of well-defined
quasiparticles is well known in one-dimensional Luttinger
liquids\cite{Hal81.1}. It is also reminiscent of the marginal Fermi liquid
behaviour in cuprate superconductors, where a large Fermi surface is seen in
the angle-resolved photoemission studies, but the scattering rate grows
linearly, rather than quadratically, in temperature\cite{Var89.1,Val99.1}. Note
also that the crossover line $T_{Hall}(B)$ does not follow the
antiferromagnetic transition line $T_N(B)$ (Fig.\,\ref{PD}a). Indeed, within
experimental resolution, the initial Hall coefficient shows no change at the
zero field N\'eel temperature of 70~mK (Fig.\,\ref{HallT}a). This behaviour
contrasts dramatically with that expected in an itinerant SDW, where changes in
the Hall coefficient should coincide with the N\'eel transition - as is indeed
observed for Cr$_{1-x}$V$_x$
(refs.\ 22 and 26). 
Thus we may discard the possibility that the observed crossover in the Hall
coefficient of YbRh$_2$Si$_2$ is due to a unit-cell doubling in a symmetry
breaking antiferromagnetic transition. Even though the crossover at
$T_{Hall}(B)$ broadens rapidly with temperature [cf.\ FWHM($T$) in the inset of
Fig.\,\ref{PD}a and width of $d\gamma(B)/dB$ in Fig.\,\ref{PD}b], so that it
can not be followed beyond about 0.5~K, the additional contribution $\Delta
R_H$ to the initial Hall coefficient (main panel of Fig.\,\ref{HallT}a) which
we attribute to fluctuations of the Fermi surface volume can be discerned up to
much higher temperatures of order 10~K. This is precisely the temperature below
which NFL behaviour is observed in thermodynamic and dynamical
properties\cite{Tro00.2,Geg02.1}. This observation makes it very tempting to
hold fluctuations of the Fermi surface volume responsible for the NFL behaviour
observed over this same temperature window. The fact that the NFL behaviour is
observed in the entire phase diagram above $T_N$ and $T^{\ast}$ (and below
10~K) can be related to the broadness of the crossover. Interestingly, also the
spin fluctuation scale $T_0$ extracted from a logarithmic fit $\Delta C_p/T
\propto ln(T_0/T)$ to the specific heat data for 0.3~K $< T <$ 10~K and the
single-ion Kondo temperature (which marks the onset of magnetic screening)
extracted from a magnetic entropy measurement are of the same order of
magnitude\cite{Tro00.2,Geg02.1}.

To summarize we observe a rapid crossover of the Hall coefficient as function
of a control parameter. By extrapolation to $T=0$ of both the Hall crossover
and the magnetic phase transition\cite{Geg02.1}, we infer that a large jump of
the Hall coefficient occurs at the QCP. We expect this new insight, made
possible primarily by the absence of superconductivity, to have broad
implications for other strongly correlated electron systems\cite{Bal03.1}.


\vspace{0.5cm}
\noindent{\large\bf Methods}

\noindent{\bf Anomalous Hall effect}

In general, the Hall effect of materials containing localized magnetic moments
is dominated at high temperatures by an anomalous Hall effect produced by the
left-right asymmetry in incoherent electron scattering processes\cite{Fer87.1}.
The {\em initial} or {\em linear-response} Hall coefficient $R_H$ (Hall
coefficient in zero-field limit) scales for many materials with the product of
electrical resistivity $\rho$ and magnetic susceptibility $\chi$,
\begin{equation}
R_H = R_0 + C\rho\chi
\label{AHE_Fer}
\end{equation}
where $R_0$ is the normal Hall coefficient and $C$ is a constant\cite{Fer87.1}.
The term $C\rho\chi$ represents the anomalous Hall effect due to {\em
intrinsic} scattering. The temperature-independent {\em extrinsic} anomalous
Hall coefficient $R_{ex}$ due to skew scattering by residual defects may be
estimated from
\begin{equation}
R_{ex} = C\rho_0\chi_0
\label{AHE_ex}
\end{equation}
where $\rho_0$ is the residual resistivity and $\chi_0$ the residual volume
magnetic susceptibility\cite{Fer87.1}. A model including crystalline electric
field effects valid in the incoherent regime\cite{Kon97.1}, on the other hand,
predicts
\begin{equation}
R_H = R_0 + R_s\chi
\label{AHE_Kon}
\end{equation}
instead of Eq.~(\ref{AHE_Fer}). Here $R_s$ is a constant and $R_s\chi$ the
anomalous Hall-effect term.

In Fig.\,\ref{HallT}a we have shown that also in YbRh$_2$Si$_2$ the
high-temperature Hall coefficient is dominated by the anomalous Hall effect.
Between 7 and 300~K (90 and 300~K), Eq.~(\ref{AHE_Fer}) [Eq.~(\ref{AHE_Kon})]
holds (cf.\ inset of Fig.\,\ref{HallT}a). The $R_0$ value obtained for both
models is $(2.4\pm 0.1)\times 10^{-10}$~m$^3$/C which corresponds, in a simple
one band model, to a charge carrier concentration of $2.6\times
10^{28}$~m$^{-3}$ (approximately 2 holes per formula unit of YbRh$_2$Si$_2$).
Considering only the magnetic contribution to $\rho$ in Eq.~(\ref{AHE_Fer})
yields similar values for $R_0$
(ref.\ 18).
Below about 1~K, where the extrapolation of the fit according to
Eq.~(\ref{AHE_Fer}) (red dashed curve in Fig.\,\ref{HallT}a) becomes
temperature independent and saturates at the value of the normal Hall
coefficient $R_0$, the intrinsic anomalous Hall effect is negligible. The
extrinsic anomalous Hall effect estimated from Eq.~(\ref{AHE_ex}) with
$\rho_0\approx 1$~$\mu\Omega$cm and $\chi_0 = 0.0035$ ($B \| c$, $T = 40$~mK)
(ref.\ 12)
is less than 4\% of $R_0$ and thus plays a negligible role. Therefore, below
about 1~K, the initial Hall coefficient of YbRh$_2$Si$_2$ is essentially free
of any anomalous contribution.

The anomalous Hall effect in {\em finite magnetic fields} may, in analogy with
Eq.~(\ref{AHE_Fer}), be estimated from
\begin{equation}
\rho_{H,a}(B) = C \rho(B) \mu_0 M(B)
\label{AHE_B}
\end{equation}
where $\rho(B)$ and $M(B)$ are the field-dependent electrical resistivity and
magnetization, respectively.

For YbRh$_2$Si$_2$, $\rho(B)$ (not shown) and $M(B)$
(ref.\ 12)
have been measured in the relevant geometry ($B \| c$, current $I \perp c$).
For the parameter $C$ we use the value extracted from the temperature
dependence of the initial Hall coefficient (inset of Fig.\,\ref{HallT}a).
$\rho_{H,a}$ is less than 20\% of $\rho_H$ at all temperatures and fields.

\vskip 1 cm


\noindent{\bf Correspondence} and requests for materials should be addressed to
S.~P. (paschen@cpfs.mpg.de).

\vspace{0.5cm}
\noindent{\bf Acknowledgements} We acknowledge discussions with P.~B\"uhler,
J.~Custers, C.~Langhammer, C.~P\'epin, A.~Rosch, A.~Schofield, M.~Vojta,
A.~Tsvelik, and F.~Weickert. Part of the work at Dresden was supported by the
Fonds der Chemischen Industrie. P.~C. and Q.~S. are supported by the National
Science Foundation. The work at Rice University was partially supported by the
Welch Foundation, and TCSAM.


\newpage



\newpage

\noindent{\bf\large Supplementary Methods 1}

\noindent{\bf Hall response from $B_2$}

In the crossed-field experiment, we establish a steady state electrical current
along $\hat{x}$, and apply two perpendicular magnetic fields: a tuning field
$B_2$ along $\hat{x}$ and a probing field $\delta B_1$ along $\hat{z}$. We
measure the transverse (Hall) voltage along $\hat{y}$. In our experiment,
$\hat{x},\hat{y},\hat{z}$ are chosen to be the principal axes and, moreover,
the crystal is isotropic in the $xy$-plane. The currents and voltages are
related by the conductivity tensor:
\begin{eqnarray}
\left( \begin{array}{ccc}
 J_x
 \\
 \\
 J_y
 \\
 \\
 J_z
 \end{array} \right)
&=&
\left( \begin{array}{ccc}
 \sigma_{\parallel} & \sigma_{xy} &  0
 \\
  & &
 \\
 -\sigma_{xy} & \sigma_{\parallel} &  \sigma_{yz}
 \\
 & &
 \\
 0 & -\sigma_{yz} &  \sigma_{\perp}
 \end{array} \right)
\times
\left( \begin{array}{ccc}
 E_x
 \\
 \\
 E_y
 \\
 \\
 E_z
 \end{array} \right).
\label{def}
\end{eqnarray}
Here we have used $\sigma_{xz}=0$, reflecting the absence of magnetic field
along $\hat{y}$, and adopted $\sigma_{\parallel}$ to denote $\sigma_{xx}$ and
$\sigma_{yy}$ (the slight difference between $\sigma_{xx}$ and $\sigma_{yy}$
induced by a finite $B_2$ can be easily incorporated in our discussion and will
not affect our conclusion below) and $\sigma_{\perp}=\sigma_{zz}$. Solving
these equations under the conditions $J_z=J_y=0$, we find
\begin{eqnarray}
R_H \equiv
\lim_{B_1\rightarrow 0} {{E_y} \over {J_x B_1}}
= \lim_{B_1\rightarrow 0}
{\sigma_{xy} \over {\sigma_{\parallel}^2 B_1}}
\times
\left [ 1+ {\sigma_{yz}^2 \over {\sigma_{\perp} \sigma_{\parallel}}}
\right ]^{-1}.
\label{rhall}
\end{eqnarray}
On the right hand side of Eq.~(\ref{rhall}), the first factor describes the
Hall constant in a gedanken setup in which the field $B_2$ does not produce any
Lorentz force while fulfilling the role of tuning the underlying state. The
second factor in principle differs from 1. For an order-of-magnitude estimate,
we can ignore the $xy$ vs $z$ anisotropy and take ${\sigma_{yz}^2 \over
{\sigma_{\perp} \sigma_{\parallel}}} \approx \left ( { \rho_H \over \rho}
\right )^2$. Given that the field $B_2$ we have applied in the crossed-field
experiment is much smaller than the largest field $B_1$ used in our
single-field experiment, $\rho_H$ here should be much smaller than $6.5 \times
10^{-10}$~$\Omega$m, the value shown in Fig.~2a of our Letter to {\em Nature}
for the temperature range $T<200$~mK. The longitudinal resistivity $\rho$ is of
the order $10^{-8}$~$\Omega$m. The second factor on the right hand side of
Eq.~(\ref{rhall}), then, differs from 1 by much less than $5\times 10^{-3}$.

To summarize, to an accuracy much better than 0.5\%, the initial slope of the
$\rho_H$ vs $B_1$ plot in the crossed-field experiment measures the
linear-response Hall coefficient of the state reached by the finite tuning
field $B_2$.

\vskip 1 cm

\noindent{\bf\large Supplementary Methods 2}

\noindent{\bf Kubo formalism for differential Hall coefficient}

The concept of a differential Hall coefficient is intimately related to
underlying current correlations inside the material. In thermal equilibrium,
the ``fluctuation dissipation theorem'' gives rise to a unique link between the
differential response of a system to a field and the fluctuations of the
variable that is coupled to the field.  These relationships are determined by
the so-called Kubo formalism~\cite{Kub57.1}. For example, the differential
magnetic susceptibility of a material is directly proportional to the two-point
spin-spin correlation function. The  orbital part of the  differential Hall
conductivity of a metal can be regarded as a current susceptibility to small
changes in the magnetic field~\cite{Vor92.1},
\begin{equation}\label{eq3}
d^{2}J_{x}
 = \delta B\delta E_{y}
\left(\frac{d\sigma_{xy}}{dB} \right)_{\rm orb},
\end{equation}
where $\delta E_{y}$ is the electric field along the $y$-axis and $J_{x}$ is
the Hall current density along the $x$-axis. In simple metals, the Lorentz
force acting on the electrons grows linearly with the applied field, giving
rise to a simple linear response $\sigma_{xy} =\frac{d\sigma_{xy}}{dB}B$. Near
a QCP, the velocities of the quasiparticles are highly sensitive to the
field-tuned ground state, so the Lorentz force changes with the tuning field
obliging us to take the field dependence of the differential Hall coefficient
into account.

Fortunately, under general conditions, the Kubo formula enables us to relate
the differential Hall response to the three-point current correlation
function~\cite{Vor92.1}
\begin{equation}\label{eq4}
Q_{\alpha\beta\gamma}(q,B) = \langle B|
J_{\alpha}(q)J_{\beta}(-\vec{q})J_{\gamma}(\omega)|B\rangle,
\end{equation}
where $q\equiv (\vec{q},\omega)$ describes the respective frequency $\omega $
and wavevector $\vec{q}$ of a probe electric and magnetic field, $J_{\alpha }$
is the current density in the $\alpha $ direction at the appropriate frequency
and wavevector and $|B\rangle $ denotes the field-tuned ground state of the
system.  All ``anomalous'' contributions to the ground-state Hall conductance
can be included into the above expression, if the $J_{\alpha }$ are taken to be
the full current operators, taking into account the momentum dependence of the
hybridization between conduction and $f$ electrons.\cite{Kon94.1} The relation
\begin{equation}\label{eq5}
Q_{\alpha \beta \gamma} (q,B)-Q_{\alpha \beta \gamma} (0,B) = \omega
(q_{\alpha }\delta _{\beta \gamma}-
q_{\beta }\delta_{\alpha \gamma})
\left( \frac{d\sigma _{xy} }{dB}\right)_{\rm orb}
\end{equation}
determines  the orbital part of  the differential  Hall conductivity. In
practice, it is more convenient to measure the differential Hall resistivity,
which is simply related to the  the differential Hall conductivity via the
relation  $\frac{d\rho_{yx}}{dB}\equiv \tilde{R}_{H}(B)= \rho^{2}
\frac{d\sigma_{xy}}{dB}$, where $\rho$ is the electrical resistivity and
$\rho_{yx} (= \rho_H)$ the Hall resistivity. In our experiments the geometry
between the probe field $\delta B_{1}$, Hall voltage and injected current is
the same in both the transverse and longitudinal field-tuning configurations,
so both  $\left [ \frac{\partial\sigma_H(B_1)}{\partial B_1} \right ]_{\rm orb}
= \left [ \tilde{R}_{H}(B_1) \right ]_{\rm orb} /\rho^2$  in the transverse and
$\sigma_{xy}(B_2,B_1)/B_1\mid_{B_1\rightarrow 0} = R_H(B_2)/\rho^2$  in the
longitudinal configuration measure the evolution of the same current correlator
through the QCP.




\newpage
\begin{figure}[ht]
\caption{\label{HallT} Temperature dependence of the Hall effect of
YbRh$_2$Si$_2$. {\bf a}, Temperature-dependent initial Hall coefficient
$R_H(T)$, obtained from the initial slope of Hall resistivity vs field
isotherms (Fig.\,\ref{HallB}a). The red curve corresponds to the red fit to the
data from the inset. $\Delta R_H$ designates the difference between the data
and the fit. The green triangles correspond to $R_H$ data obtained from the
crossed-field experiment for large values of the tuning field $B_2$
[$R^{\infty}_{H}$ values of fits to $R_H(B_2)$, cf.\ text and
Fig.\,\ref{HallB}b], suggesting that the Fermi surface volume is distinctly
larger in the field-induced paramagnetic than in the antiferromagnetic state.
Inset in {\bf a}, Initial Hall coefficient $R_H$ vs product of electrical
resistivity $\rho$ and magnetic susceptibility $\chi$ (lower axis) and vs
$\chi$ (upper axis), where temperature is an internal parameter. The full red
(black) line is a linear fit according to the anomalous Hall-effect relation
Eq.~(\ref{AHE_Fer}) [Eq.~(\ref{AHE_Kon})] to the data between 7 and 300~K (90
and 300~K), the dashed lines are the extrapolations to $T=0$. {\bf b},
Cotangent of the Hall angle $\cot\Theta_H$ ($\equiv \frac{\rho}{R_H B}$) as a
function of $T^2$, taken at $B = 1$~T. The red line (also in the inset)
corresponds to a fit, $\cot\Theta_H = C_1 + C_2 T^2$, where $C_1$ and $C_2$ are
constants. Inset in {\bf b}, Difference between data and fit (red line) of main
panel. The black line is a guide to the eye. Below 0.7~K, the data deviate
considerably from the fit.  The green squares correspond to $\cot\Theta_H$ data
obtained from the crossed-field experiment at the respective crossover fields
($B_2 = B_0$), indicating that, closer to the QCP, these deviations are even
stronger. Thus, the $\cot\Theta_H = C_1 + C_2 T^2$ behaviour appears to be a
property of the regime at elevated temperatures where quantum critical
fluctuations start to influence the physical properties, but it does not extend
over the entire temperature region down to the QCP.}
\end{figure}

\begin{figure}[ht]
\caption{\label{HallB} Magnetic field dependence of the Hall effect of
YbRh$_2$Si$_2$. {\bf a}, Single-field experiment. Typical isotherms of the Hall
resistivity $\rho_H$, corrected for its anomalous contribition $\rho_{H,a}(B)$
[Eq.~(\ref{AHE_B})], vs magnetic field $B_1=\mu_0 H_1$ ($\| c$-axis). The solid
curves represent best fits, $\int{\tilde{R}_{H}(B)dB}$ (see text), to the data.
The derivative of the fit at 75~mK is plotted on the right axis.  {\bf b},
Crossed-field experiment. Initial slope $R_H$, normalized to its value at the
crossover field $B_0$, of all measured $\rho_H$ vs $B_1$ curves as a function
of $B_2/B_0$, at 45, 65, 75, and 93~mK. The solid lines represent best fits
(see text) to the data. $R_H$ decreases by a factor of $\approx 1.5$ upon going
from the zero-field antiferromagnetic to the field-induced paramagnetic state.
In an SDW picture $R_H$ is expected to evolve as the magnetic order
parameter\cite{Col01.1}. In YbRh$_2$Si$_2$ the ordered moment at $B=0$ was
estimated to be $\approx 0.002$~$\mu_B$/Yb
(ref.\ 11).
Thus, the change in $R_H$ corresponds to a factor of $\approx 750/\mu_B$. The
corresponding change of $R_H$ by a factor of $\approx 30/\mu_B$ observed for
the SDW system Cr$_{1-x}$V$_x$
(refs.\ 22 and 26)
was already considered a giant effect, possibly connected with bandstructure
nesting effects\cite{Nor03.1}. By comparison, the effect in YbRh$_2$Si$_2$ is
about 25 times as large. Even in the absence of both experimental and
theoretical studies of the electronic bandstructure of YbRh$_2$Si$_2$ we judge
this effect far too large to be accounted for within an SDW picture. The same
agrument holds even if the second order transition observed in the measured
temperature range ($> 15$~mK) turned over into a first order one at $T <
15$~mK. The inset in {\bf b} displays $\rho_H$ vs $B_1$ curves at three
different values of the tuning field $B_2=\mu_0 H_2$~($\perp$~$c$-axis) at
65~mK. The solid lines represent best fits, as in {\bf a}. Similar data have
been obtained at the other temperatures (not shown). The sketches in {\bf a}
and {\bf b} illustrate the experimental set-up.}
\end{figure}

\begin{figure}[ht]
\caption{\label{PD} Temperature-field phase diagrams of YbRh$_2$Si$_2$. {\bf
a}, The red data points correspond to the $B_{0}$ values (crossover positions
in the Hall-effect measurements) determined from the fits to the data in
Fig.\,\ref{HallB}a (single-field experiment). Note that the horizontal bars
represent the error in the determination of $B_0$ rather than the width of the
crossover. The red dotted line denoted $T_{Hall}$ is the best linear fit to all
data up to 0.5~K. It extrapolates at zero temperature to $\approx 0.7$~T, the
critical field $B_{1c}$ for the direction parallel to the $c$-axis. The green
data points correspond to $11 B_0$ determined from the fits to the data in
Fig.\,\ref{HallB}b (crossed-field experiment). The full and dotted black curve
represent the field dependence of the N{\'e}el temperature $T_N$ and the
crossover temperature $T^{\ast}$ to a $\Delta\rho \propto T^2$ law,
respectively, as determined from iso-field $\rho(T)$ data\cite{Geg02.1}. The
latter differs qualitatively from the cross-over line determined from a scaling
analysis of both specific heat and resistivity data, yielding $T_{cross}
\propto (B-B_c)$
(ref.\ 14). 
The inset shows the full width at half maximum (FWHM) of
$d\tilde{R}_{H}(B_1)/dB_1$ in a log-log plot (red points). The red solid line,
$\propto T^a$, $a=0.5\pm0.1$, is a best fit to these data. As in the main
panel, the green dots correspond to the crossed-field experiment. For both the
main panel and the inset, the red and green data points agree within the error
bars. {\bf b}, 3D representation of the field derivative of the crossover
function $\gamma(B)$ defined in the text. The coloured curves represent
arbitrary isotherms of $d\gamma(B)/dB$, obtained using both the $B_0(T)$ fit of
{\bf a} and a power law fit to the corresponding $p(T)$ data (not shown). The
field $B$ corresponds to $B_1 \| c$ or to $11 B_2 \perp c$. The positions $B_0$
are designated by broken drop lines and the black dotted line denoted
$T_{Hall}$ in the $T-B$ plane. The antiferromagnetic phase and the region where
$\Delta\rho \propto T^2$ are marked as black and hatched areas, respectively,
in the $T-B$ plane. At the lowest temperatures, $d\gamma(B)/dB$ may be
interpreted as indicating the change of the effective carrier concentration. In
the limit $T\rightarrow 0$, $d\gamma(B)/dB$ is a $\delta$-function (dotted line
in the $T=0$ plane), separating the states of small and large Fermi surface
(FS) at $B=B_{1c}= 11 B_{2c}$.}
\end{figure}


\newpage
\begin{figure}[t]
\centerline{\includegraphics[width=140mm]{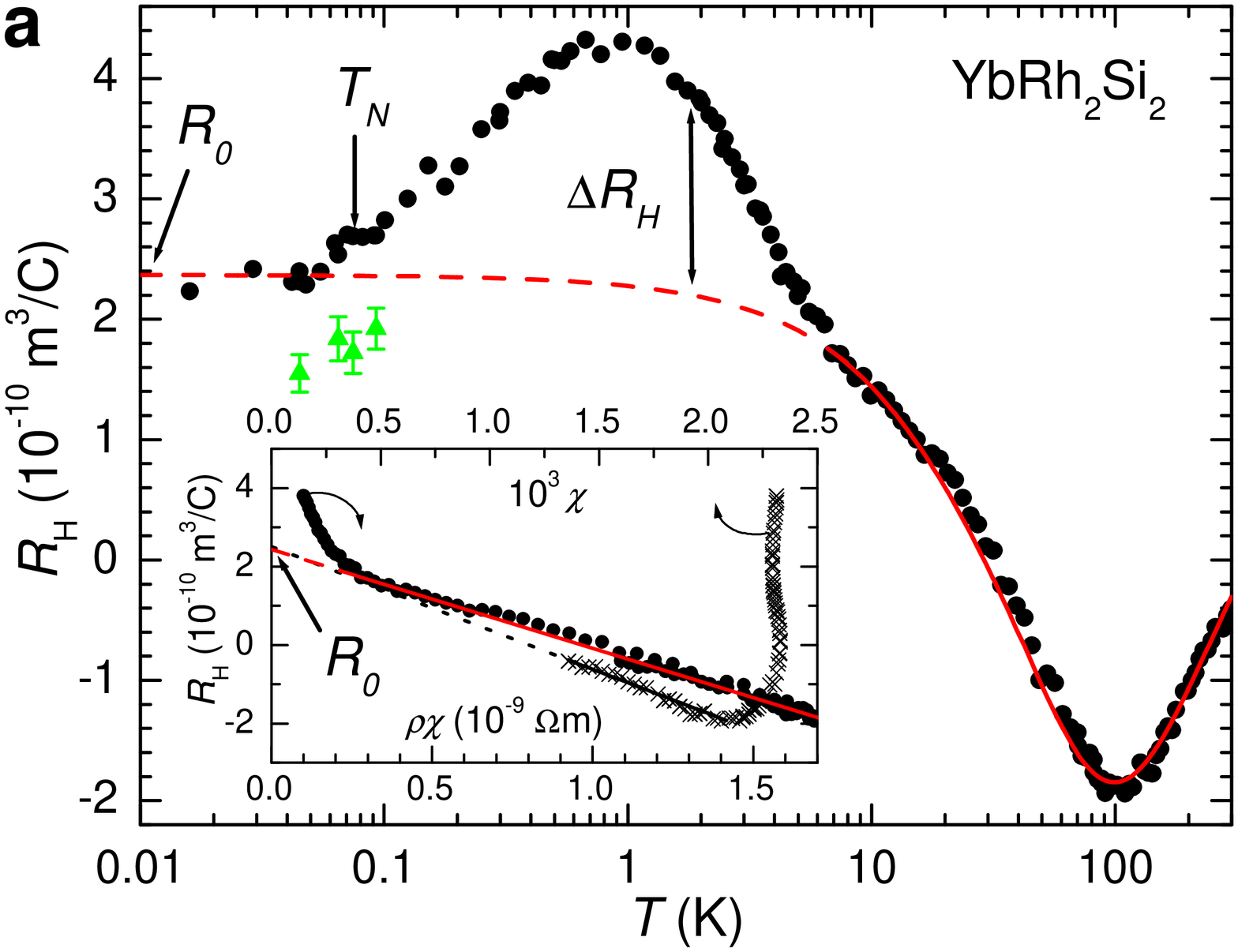}}
\vspace{0.5cm}
\centerline{\includegraphics[width=140mm]{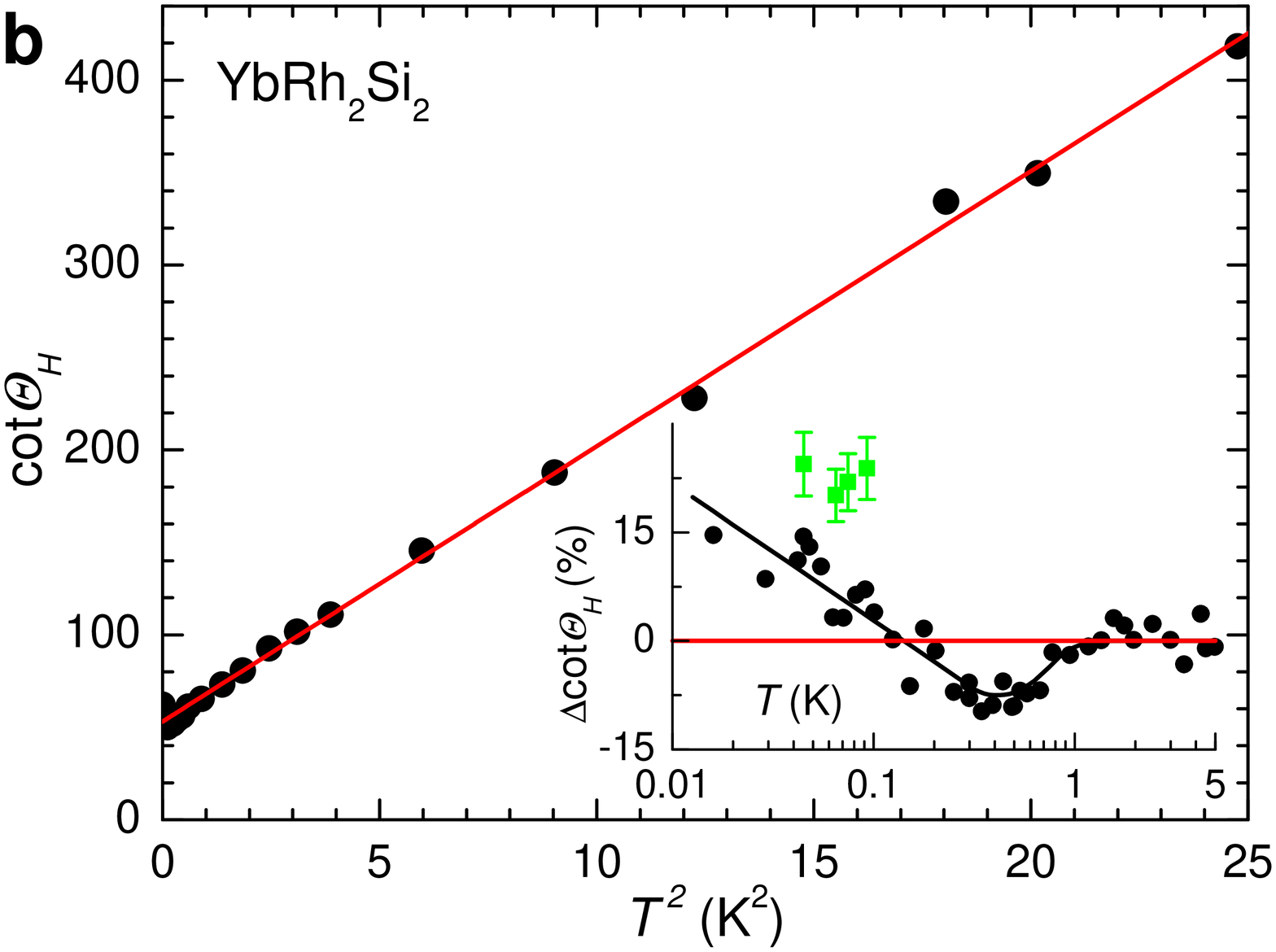}}
\vspace{0.5cm}

{\Large Figure 1}

\end{figure}

\newpage
\begin{figure}[t]
\centerline{\includegraphics[width=140mm]{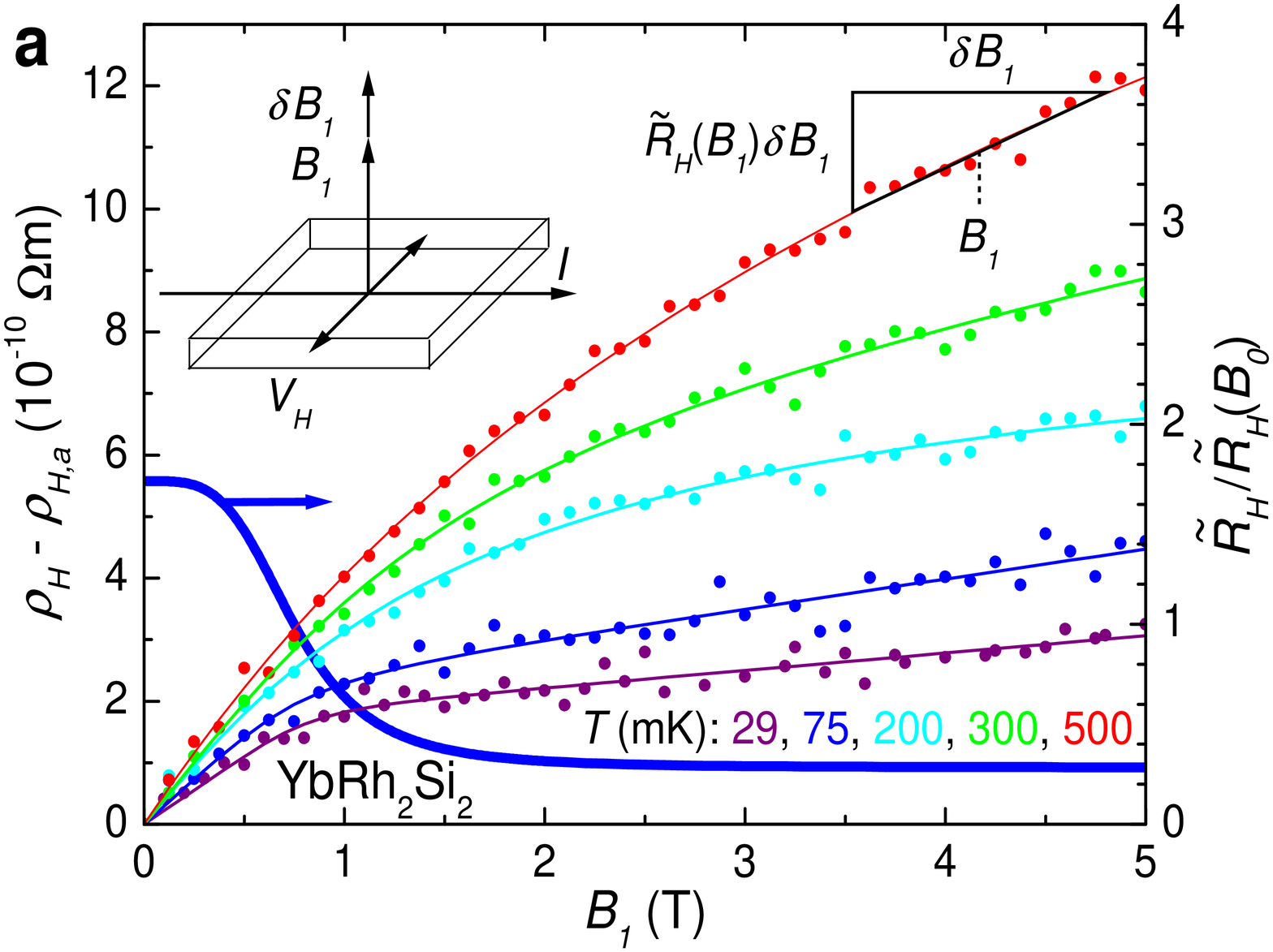}}
\vspace{0.5cm}
\centerline{\includegraphics[width=140mm]{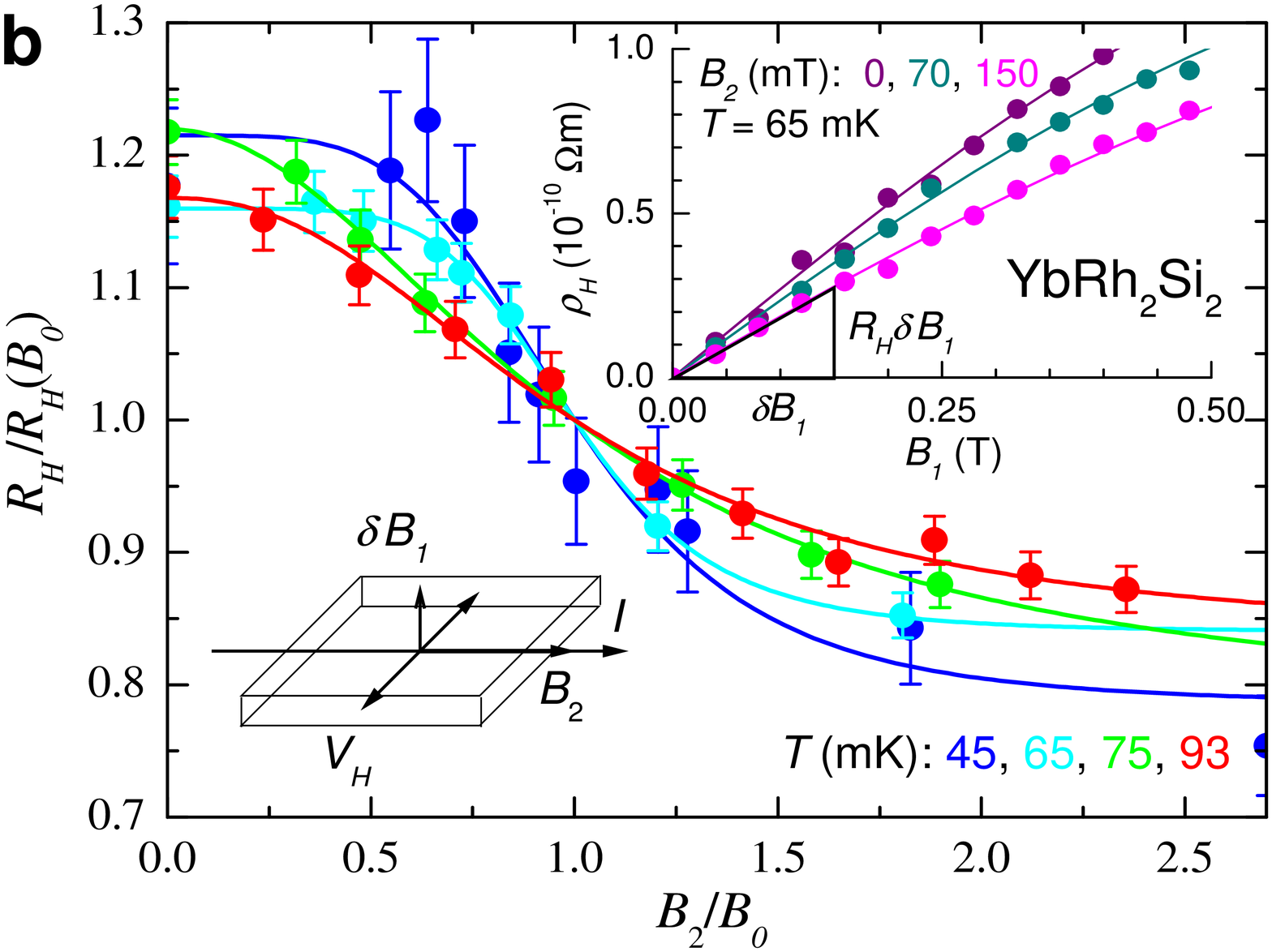}}
\vspace{0.5cm}

{\Large Figure 2}

\end{figure}

\newpage
\begin{figure}[t]
\centerline{\includegraphics[width=140mm]{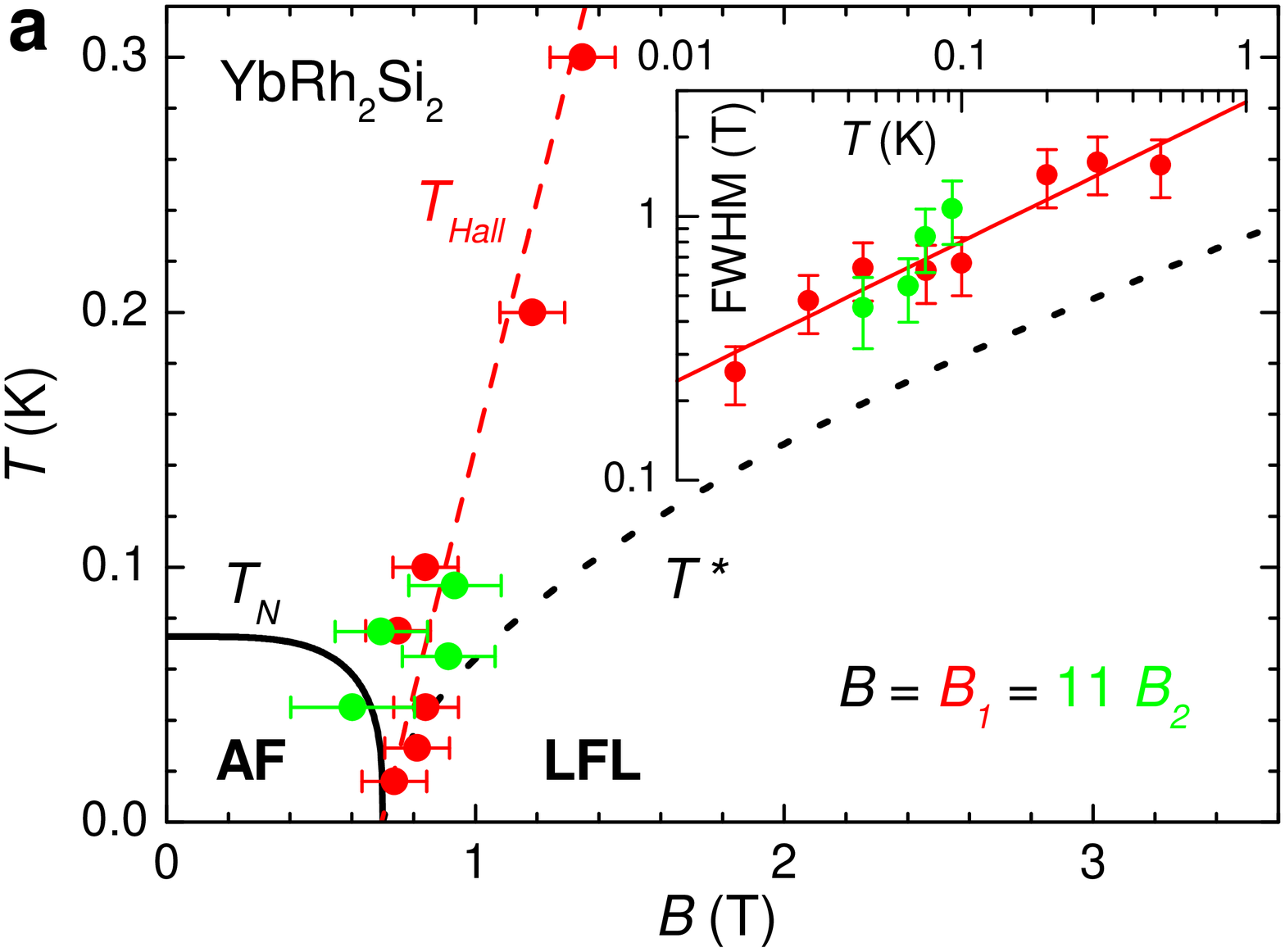}}
\vspace{0.5cm}
\centerline{\includegraphics[width=140mm]{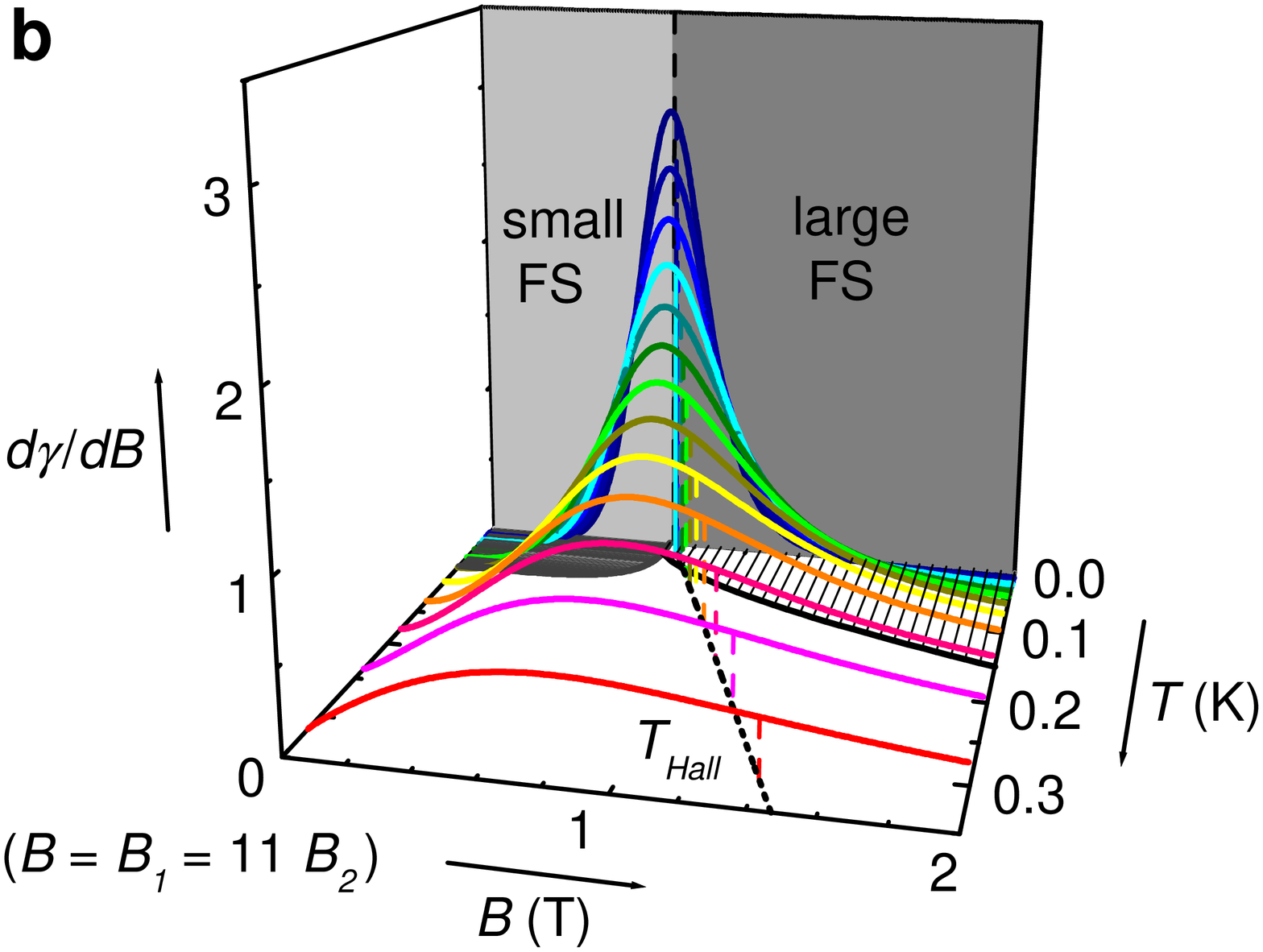}}
\vspace{0.5cm}

{\Large Figure 3}

\end{figure}



\end{document}